# Systematic investigation of influence of *n*-type doping on electron spin dephasing in CdTe

D. Sprinzl, P. Horodyská, N. Tesařová, E. Rozkotová, E. Belas, R. Grill, P. Malý, and P. Němec[*]

*Faculty of Mathematics and Physics, Charles University in Prague, Ke Karlovu 3, 121 16 Prague 2, Czech Republic*

**ABSTRACT**

We used time-resolved Kerr rotation technique to study the electron spin coherence in a comprehensive set of bulk CdTe samples with various concentrations of electrons that were supplied by n-type doping. The electron spin coherence time of 40 ps was observed at temperature of 7 K in *p*-type CdTe and in *n*-type CdTe with a low concentration of electrons. The increase of the concentration of electrons leads to a substantial prolongation of the spin coherence time, which can be as long as 2.5 ns at 7 K in optimally doped samples, and to a modification of the *g* factor of electrons. The influence of the concentration of electrons is the most pronounced at low temperatures but it has a sizable effect also at room temperature. The optimal concentration of electrons to achieve the longest spin coherence time is 17-times higher in CdTe than in GaAs and the maximal low-temperature value of the spin coherence time in CdTe is 70-times shorter than the corresponding value in GaAs. Our data can help in cross-checking the predictions of various theoretical models that were suggested in literature as an explanation of the observed non-monotonous doping dependence of the electron spin coherence time in GaAs.

## I. INTRODUCTION

The utilization of the electron spin, in addition to its charge, is a heart of the emerging new branch of electronics – spintronics. The dephasing time of a spin coherence, which directly determines the survival of information encoded using the spin of carriers, is one of the most important material parameters for spintronics and quantum computation. Therefore, the observed suppression of the spin dephasing in *n*-type GaAs[1] motivated a rather intensive research in this field. And even though the influence of *n*-type doping on the spin relaxation was experimentally observed also in several other bulk materials (namely, ZnSe[2], GaN[3,4], InSb[5,6], InAs[7], and ZnO[8]) the topic remains controversial. Up to now, a *systematic* study of the dependence of the electron spin dephasing time on the concentration of electrons (*n*) was reported only in GaAs[9] and, very recently, in InSb[6] (in other materials, typically, only two or three different doping levels were studied). And the obtained results are quite different in these two materials. In GaAs the increase of *n* leads to a prolongation of the spin dephasing time for *n* up to $3 \times 10^{15}$ cm$^{-3}$ that is followed by its decrease for higher *n* (Ref. 9). On the other hand, in InSb the spin relaxation time was reported to *decrease* with *n* (Ref. 6), which, in fact, contradicts the earlier results[5] reported for this material. Also from the theoretical point of view this issue is not fully understood. Namely, two different spin relaxation mechanisms (SRM) were stated to be dominant in *n*-type semiconductors at low temperatures – Elliot-Yafet (EY)[10,11] and D'yakonov-Perel (DP)[12,13]. Alternatively, the hyperfine interaction and/or a change of dominant SRM with *n* were also considered.[r4,9,14] In addition, a considerably different mechanisms were used to explain the observed non-monotonous dependence of the spin dephasing time on *n* in GaAs.[9, 12, 13] In this paper, we would like to contribute to this topic by providing another model material where the dependence of the electron spin dephasing time on *n* was measured systematically. We selected CdTe for this research because it has the same crystal structure and nearly the same band gap as GaAs but rather different material parameters (e.g., effective masses, dielectric constants and spin-orbit interaction). Moreover, the comparison of the recently published data in *n*-type doped CdTe quantum wells[15] with our data measured in the corresponding bulk material can help in the understanding of the role of the quantum confinement in the spin dynamics. Prior to this work only room temperature data in intentionally undoped bulk CdTe were reported[16-18].

## II. EXPERIMENTAL

CdTe bulk single crystals were prepared by a vertical gradient freeze method. All undoped as-grown samples were *p*-type (with a hole concentration $p \approx 10^{14} - 10^{16}$ cm$^{-3}$) due to an unintentional doping by foreign acceptors. *N*-type CdTe single crystals were prepared by the intentional donor doping (using indium). The electron concentration *n* can be tuned in the interval $10^7 - 10^{18}$ cm$^{-3}$ using a post-growth annealing of In-doped crystals in the various Cd overpressures.[19] The highest electron concentration, which is similar to the donor doping level, was obtained by the annealing at the Cd-saturated overpressure. Crystals with a lower electron concentration can be prepared by the annealing in a lower Cd overpressure. We studied eight *n*-type samples with *n* spanning from $1.5 \times 10^{13}$ cm$^{-3}$ to $3.2 \times 10^{17}$ cm$^{-3}$ and also, as a reference, the as-grown *p*-type sample with $p \approx 10^{16}$ cm$^{-3}$ (the carrier concentration was determined at room temperature by the Hall effect measurement).

The dephasing of electron spin coherence was studied by the time-resolved Kerr rotation (KR) technique using a femtosecond Ti-sapphire laser (Tsunami, Spectra Physics). Spin-polarized electrons were optically injected by laser pulses with a duration of 80 fs and a repetition rate of 82 MHz, which were spectrally tuned to match the band gap energy of CdTe (i.e., during the measurements of the temperature dependences the wavelength of laser pulses was tuned to follow the shrinkage of a band gap of CdTe with the sample temperature[20]). The pump laser polarization was modulated by a photoelastic modulator from left circular to right circular at 50 kHz that eliminated the buildup of a nuclear spin polarization via hyperfine interaction. The energy fluence of the pump pulses was about 1 µJ.cm$^{-2}$, which corresponds to a concentration of photoexcited carriers of about $5 \times 10^{16}$ cm$^{-3}$, and the probe pulses were always at least 10 times weaker. The sample was mounted in a closed-cycle He cryostat placed between the poles of an electromagnet, which created a transverse magnetic field up to $\approx 0.7$ T.

## III. RESULTS AND DISCUSSION

In Fig. 1 (a) we show the KR signals measured in CdTe crystals with different concentration of electrons as a function of the time delay between pump and probe pulses ($\Delta t$). The signals can be fitted by a function

$$KR(\Delta t) = \sum_{i=1}^{2} A_i \exp(-\Delta t / t_i) \cos(\omega_L \Delta t), \qquad (1)$$

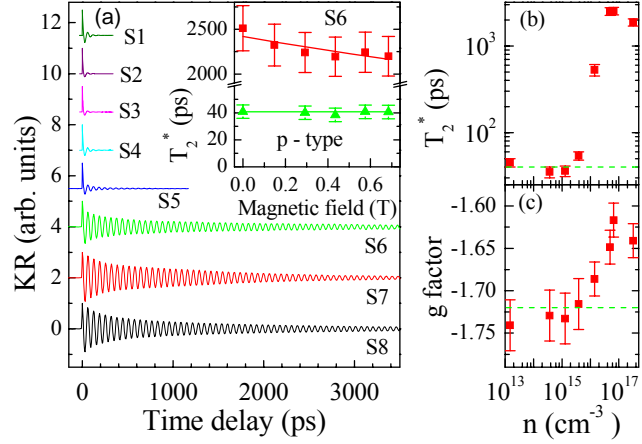

Fig. 1. (a) Time-resolved KR signal measured in *n*-CdTe with different concentrations of electrons (in cm$^{-3}$) - sample S1: $1.5 \times 10^{13}$, S2: $3.7 \times 10^{14}$, S3: $1.3 \times 10^{15}$, S4: $3.9 \times 10^{15}$, S5: $1.4 \times 10^{16}$, S6: $4.9 \times 10^{16}$, S7: $6.6 \times 10^{16}$, S8: $3.2 \times 10^{17}$; the data are normalized and offset for clarity. The measurement was done in a magnetic field of 0.685 T at a temperature of 7 K. Inset: Dependence of the spin coherence time $T_2^*$ on the transverse magnetic field for the sample S6 (with an optimal concentration of electrons) and the reference *p*-CdTe (points). The upper line depicts the inhomogeneous dephasing $1/T_2^* \approx 1/T_2^*(0) + \Delta g \mu_B B / 2\hbar$ (Ref. 22) in S6 (with a Gaussian distribution of g factors $\Delta g = 0.013$) and the horizontal lower line illustrates the independence of $T_2^*$ on the magnetic field for the *p*-type sample. (b) and (c) Dependence of the spin coherence time $T_2^*$ and the *g* factor, respectively, on the concentration of electrons *n* for a magnetic field of 0 T and a temperature of 7 K. The corresponding values in the reference *p*-CdTe sample are shown as horizontal dashed lines.

where $A_i$ and $t_i$ describe the amplitude and the decay time of the signal envelope, respectively, and $\omega_L$ is the Larmor frequency, which is a direct measure of the electron *g* factor, $g = (\hbar \omega_L)/(\mu_B B)$ ($\mu_B$ is the Bohr magneton and *B* is the magnetic field). In the investigated samples the signal envelope could not be described well by a simple mono-exponential decay. Instead, a double-exponential decay seems to be more appropriate. Similar double-exponential (or multi-exponential) decay was observed also in other materials[3, 4, 15, 21] and we attribute the shorter and the longer time constants to the lifetime of photo-injected carriers and to the electron transverse spin coherence time $T_2^*$, respectively.[4] In Fig. 1 (b) we show a dependence of the zero field values of $T_2^*$ on the concentration of electrons *n*. For small *n*, the obtained values of $T_2^*$ are the same as that measured in the reference *p*-type sample, which is probably connected with the influence of photoexcited carriers with a concentration of about $5 \times 10^{16}$ cm$^{-3}$. When *n* exceeds the value of $\approx 5 \times 10^{15}$ cm$^{-3}$ there is a rather substantial prolongation of the spin coherence time $T_2^*$ (up to $\approx 2.5$ ns for the samples with the optimal value of the concentration of electrons $n^{opt} \approx 5 \times 10^{16}$ cm$^{-3}$) that is followed by a decrease of $T_2^*$ for $n \geq 10^{17}$ cm$^{-3}$. This prolongation, which is up to 60-times in samples with $n \approx n^{opt}$ with respect to $T_2^*$ value observed in the reference *p*-type sample [the measured value $T_2^* \approx 40$ ps is schematically shown as a horizontal dashed line in Fig. 1 (b)], is accompanied also by a change of the magnetic field dependence of $T_2^*$. In *p*-type sample, $T_2^*$ does not depend on the magnetic field but in the sample with $n^{opt}$ the values of $T_2^*$ decrease

arXiv 1001.0869

with a magnetic field – see Inset in Fig. 1 (a). Similar doping-induced change of magnetic field dependence of $T_2^*$ was observed also in other semiconductors.[1, 3, 8, 15, 22] This field dependence of $T_2^*$ (i.e., the time constant that characterizes the decay of the oscillation envelope) in the sample with $n^{opt}$ can be a consequence of a "dephasing" of precession angles within the excited spin population that arises from a spread in the electron $g$ factor.[22] Alternatively, the decrease of $T_2^*$ with a magnetic field can be explained by an additional magnetic field-dependent SRM.[3]

Our data further reveal that in CdTe there is also a systematic dependence of the $g$ factors on the concentration of electrons. In fact, from the KR experiment we can determine only the magnitude of the $g$ factor but not its sign. Nevertheless, we can obtain the sign of the $g$ factor from a comparison with the previously reported[23] low-temperature value for bulk CdTe $g = -1.65 \pm 0.03$. The measured dependence of the electron $g$ factor on $n$ is shown in Fig. 1 (c). The obtained doping dependence of the $g$ factors is quite similar to that of $T_2^*$ - for $n < 5 \times 10^{15}$ cm$^{-3}$ the values of the $g$ factors in $n$-type samples are the same as that in the $p$-type sample and above this concentration there is a reduction of the $g$ factor magnitude. The similar values of the $g$ factors in $p$-type and $n$-type samples clearly shows that we measure the spin relaxation of electrons also in the $p$-type sample (because the electron and the hole g factors are known to be quite different). Up to now, the doping dependence of $g$ factors was not studied systematically in $n$-type semiconductors. We will come back to the discussion of this effect later on.

The measured non-monotonous dependence of the spin coherence time on the concentration of electrons [Fig. 1 (b)] is quite different from that measured recently in InSb[6] but it is rather similar to that observed in GaAs (see Fig. 3 in Ref. 9). The major similarity between GaAs and CdTe is that in both materials there exist an optimal concentration of electrons $n^{opt}$ where $T_2^*$ is the largest. Nevertheless, there are significant differences also between GaAs and CdTe: in GaAs the largest values of $T_2^*$ ($\approx 180$ ns) are obtained for $n^{opt} \approx 0.3 \times 10^{16}$ cm$^{-3}$, in CdTe the largest values of $T_2^*$ ($\approx 2.5$ ns) are obtained for $n^{opt} \approx 5 \times 10^{16}$ cm$^{-3}$. We note that in samples with a low concentration of electrons the value of $T_2^*$ is rather similar in both materials – 65 ps in GaAs[24] and 40 ps in CdTe. The exact physical origin of the peak in the doping dependence of $T_2^*$ in GaAs is still an opened question – this density was even called a 'magic' electron density.[22] In Ref. 9 this maximum, which was observed for $n$ close to the metal-to-insulator transition (MIT) in GaAs, was assigned to the crossover between relaxation mechanisms originating from the hyperfine interaction of localized electrons with lattice nuclei and from the spin-orbit interaction of free electrons in the metallic regime. This explanation was adopted also in Ref. 14 where the bias-dependent electron spin lifetimes were measured by the Henle effect using the cw laser. On the other hand, it was argued in Ref. 13 that this peak can be explained solely by a breakdown of the motional narrowing in DP SRM – i.e., that MIT does not have to be considered. Similarly, in Ref. 12 this non-monotonous dependence of $T_2^*$ on $n$ was explained by DP SRM only - $T_2^*$ increases with $n$ in the nondegenerate regime (i.e., for low values of $n$) due to a decrease of the momentum scattering time but it decreases in the degenerate regime (i.e., for high values of $n$) due to an enhancement of the inhomogeneous broadening. Our results clearly revealed that a considerably higher value of $n^{opt}$ is required in CdTe than in GaAs for a maximal suppression of the electron spin dephasing. One possibility is that it is connected with the MIT. In general, MIT occurs above the critical Mott concentration $n_c$ that is given by $n_c^{1/3} a_B = 0.25$, where $a_B$ is the exciton Bohr radius.[25] In GaAs $n_c \approx 2 \times 10^{16}$ cm$^{-3}$ (Ref. 9), so, $n^{opt} \approx 0.15\, n_c$. For uncompensated CdTe MIT occurs for $n_c \approx 15 \times 10^{16}$ cm$^{-3}$ but, for some degree of compensation in the samples, $n_c$ could be as high as $90 \times 10^{16}$ cm$^{-3}$ (Ref. 26). Taking this uncertainty in the determination of $n_c$ into account, we obtained for CdTe the relation $n^{opt} \approx$ (0.05 - 0.3) $n_c$, which is quite in line with the results observed in GaAs. These results seems to indicate that there might be indeed a connection between the MIT and the peak in the doping dependence of $T_2^*$, as suggested in the earlier reports.[9, 14] However, because we cannot exclude the possibility that this apparent correlation between $n^{opt}$ and $n_c$ is just a coincidence, for its verification it would be necessary to measure the systematic doping dependence of $T_2^*$ yet in another material.

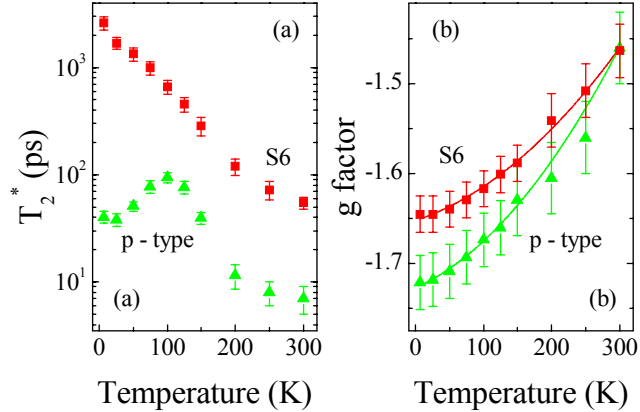

Fig. 2. Temperature dependence of the spin coherence time $T_2^*$ (a) and the g factor (b) measured in the n-CdTe sample S6 and the reference p-CdTe sample for a magnetic field of 0 T. The lines in (b) are fits by the function $g(T) = a + bT + cT^2$ (Ref. 23), where a = -1.65, b = 2.5 × 10$^{-4}$, c = 1.3 × 10$^{-6}$ for the sample S6, and a = -1.73, b = 3.4 × 10$^{-4}$, c = 1.9 × 10$^{-6}$ for the p-type sample.

The dominant SRM for samples with $n^{opt}$ is still an opened question even for GaAs, which is the most thoroughly investigated semiconductor.[9-13] For example, it was suggested in Ref. 10 that EY SRM is dominant in strongly n-doped semiconductors. On the contrary, in Ref. 12, where the electron spin relaxation was investigated from a fully microscopic kinetic spin Bloch equation approach, it was concluded that EY SRM is less important than DP SRM. On the other hand, in Ref. 11 it was argued that at low temperatures DP SRM is not applicable below MIT, where the electrons are strongly localized. In any case it should be stressed that in Ref. 10 rather simple approximate formulas were used while in Refs. 12 and 13 much more complex and accurate theoretical description was provided. Our major aim in this paper is to provide an experimental data for a second model material where the microscopic calculations can be cross-checked with the experimental results that could help in solving the ongoing controversy in the field. In Fig. 2 the temperature dependence of the spin dephasing is shown for the p-type doped and the optimally n-type doped CdTe samples. There is a clear difference between these two samples – in the n-type sample $T_2^*$ decreases monotonously with increasing temperature while in the p-type sample $T_2^*$ increases for temperatures up to 100 K and then starts to decrease. Even though the precise determination of the dominant SRM in CdTe at various temperatures and concentrations of electrons is out of the scope of this paper, we can discuss the role of various SRMs in CdTe qualitatively. First, we address a possible role of the hyperfine interaction for a spin relaxation of the donor-bound electrons. The averaging of the nuclear field from the atoms, which are within the localization volume of the impurity bound electron, leads to the dephasing time that is proportional to the number of the atoms and inversely proportional to the strength of the hyperfine interaction. [9, 27, 28] In GaAs 100% of nuclei have spin 3/2 (see Table 1 in Ref. 28) and, therefore, the hyperfine constant, which characterizes the strength of the hyperfine interaction and thus the precession frequency of the electron spin in the nuclear field, is rather large: $A_{GaAs}$ = 90 μeV (Ref. 29). In CdTe only a fraction of the nuclei have a magnetic moment (8% of *Te* and 25% of *Cd* ions) and these have a spin 1/2 (see Table 1 in Ref. 28). As a result, the spin interaction with the nuclei is rather weak in CdTe. If we suppose that the hyperfine interaction is the dominant SRM in GaAs for $n^{opt}$ where $T_2^* \approx 180$ ns (Ref. 9) and if we take, for simplicity, the value $A_{Cd}$ = 12 μeV as the hyperfine constant of CdTe (Ref. 27) and if we assume that the localization volume of the bound electron in CdTe is similar to that in GaAs, the electron dephasing time due to hyperfine interaction should be approximately 4-times *longer* in CdTe than in GaAs. Even though this is only a rough estimate, as the actual localization volume of the electrons in not known for CdTe, it suggests that the hyperfine interaction is rather unlikely to be the dominant SRM in *n*-type CdTe as the measured maximal value of $T_2^*$ in CdTe is 70-times *shorter* than the corresponding value in GaAs. The Bir-Aronov-Pikus (BAP) SRM, in which electrons exchange their spins with holes, is also ineffective due to the lack of holes in *n*-type semiconductors. Consequently, the dominant SRM is probably the DP or EY SRM (or their combination). At this time we are not able to evaluate their relative importance - for this it is necessary to perform a fully microscopic calculation of EY SRM.[12] We just want to mention that the smaller value of $T_2^*$ measured in optimally doped CdTe compared to that observed in GaAs might be connected with the stronger spin-orbit interaction in CdTe compared to that in GaAs (the spin-orbit splitting $\Delta_0$ = 0.80 eV in CdTe and 0.341 eV in GaAs)[30].

In Fig. 2 (b) we show the corresponding temperature dependence of *g* factors. While at low temperatures the g factors in the *n*-type and *p*-type samples are rather different [cf. Fig. 1 (c)], at 300 K they are nearly identical. Overall, the temperature dependence of the *g* factors is quite similar to those measured in GaAs[23, 31, 32] and CdTe[23]. The *g* factors are affected by the lattice temperature by two competing processes: As the temperature increases the band gap energy decreases, which is making the g factors more negative.[31, 32] On the other hand, with the increasing temperature the electrons populate higher Landau levels and the band's nonparabolicity makes the g factors more positive.[31] The second effect could be also responsible for the observed doping dependence of the g factors [see Fig. 1 (c)] – as the concentration of electrons increases they populate higher Landau levels that in turn makes their average *g* factor less negative.

## IV. CONCLUSIONS

In conclusion, we performed a detailed measurement of the electron spin dephasing in a comprehensive set of *n*-CdTe bulk crystals with various concentrations of electrons (from 1.5 × 10$^{13}$ cm$^{-3}$ to 3.2 × 10$^{17}$ cm$^{-3}$) and also, as a reference, in the *p*-CdTe sample with a concentration of holes ≈ 10$^{16}$ cm$^{-3}$. The major goal of this Brief Report was to provide the experimental data for a second bulk model material where the suppression of the spin dephasing time of electrons by the *n*-type doping was studied systematically. In particular, we showed that in the samples with a low concentration of electrons $T_2^*$ is comparable in CdTe and GaAs. For the optimal concentration of electrons, which is 17-times higher in CdTe than in GaAs, $T_2^*$ is significantly prolonged in both materials but the maximal value of $T_2^*$ is 70-times shorter in CdTe than in GaAs. We believe that our CdTe data can help in cross-checking the predictions of various theoretical models that were suggested as an explanation of the observed non-monotonous doping dependence of $T_2^*$ in GaAs.

## ACKNOWLEDGEMENTS

This work was supported by Ministry of Education of the Czech Republic (research centre LC510 and the research plan MSM0021620834), Grant Agency of the Czech Republic (grant no. 202/09/H041) and by grant no. SVV-2010-261306 of the Charles University in Prague.